# Electromagnetic turbulence in EAST plasmas with internal transport barrier

**Yuehao Ma[1], Pengfei Liu[2], Jian Bao[2], Zhihong Lin[3] and Huishan Cai[1,*]**

[1] CAS Key Laboratory of Frontier Physics in Controlled Nuclear Fusion, School of Nuclear Sciences and Technology, University of Science and Technology of China, Hefei 230026, China
[2] Institute of Physics, Chinese Academy of Sciences, Beijing 100190, China
[3] Department of Physics and Astronomy, University of California, Irvine, CA 92697, United States of America

E-mail: hscai@mail.ustc.edu.cn

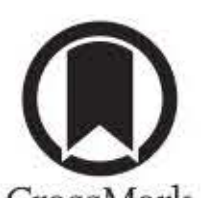

**Abstract**

In this study, global nonlinear electromagnetic gyrokinetic simulations are conducted to investigate turbulence in the internal transport barrier region of the EAST discharge with weakly reversed magnetic shear. Linear simulations reveal two dominant ion temperature gradient (ITG) modes: a higher frequency mode at the $q = 1$ surface, which dominates in the electrostatic limit, and a lower frequency mode near the $q_{\mathrm{min}}$ surface, which prevails under the experimental $\beta$ (the ratio of plasma pressure to magnetic pressure). Therefore, electromagnetic effects play a important role in stabilizing ITG modes, and in causing the transition between the most unstable mode at different radial positions. The linear growth rate of the unstable mode in the electrostatic limit is approximately 1.25 times higher than that of the dominant mode in the electromagnetic case. However, in the electromagnetic nonlinear regime, the thermal ion heat conductivity is reduced by at least a factor of 4. This reduction primarily results from nonlinear electromagnetic effects enhancing the shearing effect of zonal flows, thereby further suppressing microturbulence. It is emphasized that the electromagnetic effect on ITG with weak magnetic shear should be included to accurately calculate the turbulent transport.

(Some figures may appear in colour only in the online journal)

## 1. Introduction

Microturbulence plays a critical role in the anomalous transport of heat and particles in magnetically confined plasmas [1–3]. In general, the ion temperature gradient (ITG) turbulence primarily drives anomalous ion heat transport in the tokamak plasmas [4, 5]. The kinetic ballooning mode (KBM) [6–8], an electromagnetic instability that is destabilized at high plasma $\beta$, where $\beta$ represents the ratio of plasma pressure to magnetic pressure. Suppression of these turbulence reduces core plasma transport and facilitates the formation of internal transport barrier (ITB), which are important for achieving advanced tokamak operational scenarios [9, 10].

Many factors contribute to the stabilization of turbulence in the core region of tokamak plasmas. For example, magnetic geometry effects such as reversed magnetic shear or the Shafranov shift, can modify the linear growth rates of instabilities [11–13]. Additionally, the parallel ion transit term provides further stabilization to toroidal ITG modes, particularly in core regions with small safety factor [14, 15]. Gyrokinetic simulations and experiments have demonstrated that energetic particles (EPs) generated by auxiliary heating

* Author to whom any correspondence should be addressed.

schemes can beneficially impact ITG turbulence transport [16–19]. Many of these stabilizing factors are captured in electrostatic case; however, electromagnetic global gyrokinetic simulations are essential for accurately studying turbulent transport, as finite $\beta$ can stabilize turbulence through magnetic field line bending as plasma beta increases [5, 20–24].

Recently, simulation studies of electromagnetic turbulence and experiments have been conducted in devices such as DIII-D, JET, ASDEX Upgrade, KSTAR and HL-2A [25–32]. Global gyrokinetic simulations find the electromagnetic effects greatly reduce the ITG induced thermal ion heat transport by a factor of 10 in DIII-D plasmas [25]. It has been found that energy transport driven by ITG turbulence does not decrease with increasing plasma $\beta$ [33, 34]. Moreover, the electromagnetic stabilization becomes notably effective under conditions of low global magnetic shear [17, 29, 33, 35]. The formation of ITB in the EAST often benefits from configurations characterized by weakly reversed magnetic shear [36, 37], where electromagnetic effects likely play a significant role. Therefore, it is necessary to conduct gyrokinetic simulations that incorporate electromagnetic effects in the EAST ITB plasmas. In this work, we performed global simulations of electromagnetic turbulence using the first-principle gyrokinetic toroidal code (GTC) [38] based on EAST discharge [37]. Electromagnetic effects significantly suppress the higher frequency ITG mode that dominates in the electrostatic case. Hence, a lower frequency ITG near the $q_{\min}$ surface becomes dominant at the experimental $\beta$, and the ion heat conductivity is reduced by a factor of 4 relative to the electrostatic limit. The presence of EPs results in slightly reduced linear growth rates of ITG instability and induced zonal flows.

This paper is organized as follows: The physics model and simulation parameters are presented in sections 2.1 and 2.2, respectively. Linear simulation results are discussed in section 2.3, while turbulence nonlinear saturation and transport are analyzed in section 2.4. Finally, the summary and discussions are provided in section 3.

## 2. GTC simulation results of EAST ITB plasmas

### 2.1. Physical model in GTC

The GTC is a particle-in-cell (PIC) code developed to simulate plasma behavior and turbulence transport in fusion reactors. Over time, GTC has incorporated several key physical models, including the kinetic electron response [39], electromagnetic modeling [40], equilibrium current [41], and compressional magnetic perturbations [42]. The electromagnetic capability of GTC is implemented using a fluid-kinetic hybrid electron model [40, 43]. In this model, the electron response is separated into adiabatic and nonadiabatic components, with the adiabatic part solved by massless fluid response, and the nonadiabatic part solved using the drift-kinetic equation. The fluid-kinetic hybrid model has been successfully verified for microturbulence [41, 44], Alfvén eigenmodes [45], and current or pressure-driven MHD modes [46, 47], further establishing GTC as a powerful tool for simulating multi-scale [48] plasma physics in fusion devices.

### 2.2. Simulation parameters

The equilibrium of EAST tokamak discharge #93890 at 5000 ms [37], characterized by a weakly reversed magnetic shear $q$ profile and ITBs in both ion density and temperature profiles, was previously detailed in our electrostatic turbulence study [49, 50]. Building on this foundation, our current work advances into electromagnetic simulations, with figure 1 showing the plasma equilibrium. The major radius on the magnetic axis is $R_0 = 1.91$ m, and on-axis magnetic field amplitude is $B_0 = 1.49$ T. Notably, regions exhibiting ITBs are identified approximately within $r/a < 0.4$. As shown in figure 1(*b*), the density profile of EPs was computed using the NUBEAM module integrated into the ONETWO transport solver [51], with the neutral beam injection configured at an energy of approximately 50 keV. A power balance analysis has also been carried out using the ONETWO code, through which the ion thermal transport can be calculated [37]. The temperature profile of the energetic ions, equivalently calculated using a slowing-down distribution, is assumed to be radially uniform with $T_f = 15$ keV. Furthermore, this experimental scenario based on the low $q_{95}$ operation regime exhibits a high plasma beta, as shown in figure 1(*d*). As beta increases, electromagnetic effects, including the finite beta effect, become more significant and cannot be neglected when analyzing plasma transport. Therefore, electromagnetic gyrokinetic simulations are essential to understand the role of these effects in determining the transport properties of the EAST ITB plasma.

### 2.3. Linear simulation results

In our previous electrostatic studies [49], we identified the ITG mode as the dominant instability in the ITB region, localized at the maximum ITG and the $q = 1$ surface, with simulations incorporating both adiabatic and kinetic electron models. In this work, all electromagnetic simulations employ the kinetic electron model, incorporating the kinetic effects of the electron response. The characteristics of the ITG mode vary significantly across radial positions due to differences in equilibrium profiles and magnetic geometry. Thus, by strategically restricting the simulation domain to localized regions, we can identify both the dominant and subdominant ITG modes. Figure 2(*a*) displays the experimental $\beta_i$ and $q$ profiles, with $\beta_{i0}$ on axis reaching approximately 2.02%. Two vertical dashed lines indicate the positions of two dominant ITG modes. One higher frequency ITG mode is located at the $q = 1$ surface with $r_1/a = 0.25$, where $R_0/L_{Ti}$ is maximal and the experimental $\beta_{i,r_1}$ is 1.1%, prevailing in the electrostatic limit [49]. However, a lower frequency ITG mode near the $q_{\min}$ surface at $r_2/a = 0.35$ is dominant at the experimental plasma beta. These modes are discussed in detail below.

The properties of instability and their dependence on $\beta_i$ are investigated by changing the electron density, while the gradient of electron density is kept unchanged. This results in a vertical shift of the $\beta_i$ profile. In the simulation results shown in figure 2(*b*), the negative mode frequency indicates that the mode propagates in the direction of the ion diamagnetic drift. In the electrostatic case, the ITG mode at the $q = 1$ surface is the most unstable mode in the ITB region [49], with the mode structure of $\delta\phi$ shown in figure 3($a$). However, as the $\beta_{i0}$ increases, the growth rate of this mode decreases significantly, with the corresponding mode structure displayed in figure 3($c$). In contrast, the lower frequency ITG mode near the $q_{\rm min}$ surface, with the mode structure shown in figures 3($b$) and ($d$), exhibits a smaller reduction in growth rate as $\beta_{i0}$ increases. Once the $\beta_{i0}$ exceeds 0.5%, this higher frequency ITG mode is no longer dominant. Thus, at the experimental plasma beta, the mode near $q_{\rm min}$ becomes the dominant instability in the ITB region. If only electrostatic simulations are considered, the transition between the most unstable mode at the different radial position may be missed. This phenomenon can be understood from the critical condition for electromagnetic stabilization of the ITG instability, typically characterized by $\beta_{\mathrm{crit}} \sim 1/q^2 L_{T_i}/R_0$ [5, 24]. At the $q = 1$ surface, the inverse temperature scale length $R_0/L_{T_i}$ (shown in figure 1(*c*)) tends to be larger compared to its value at the $q_{\min}$ surface, resulting in a correspondingly smaller critical $\beta$. Hence, finite $\beta$ effects can more effectively stabilize the ITG mode at locations with a larger temperature gradient, requiring only a lower $\beta_i$ to achieve a substantial reduction in the linear growth rate. On the other hand, in the cyclone base case (CBC) [52], the growth rate of ITG modes slightly decreases with increasing $\beta_i$ with the normal magnetic shear $q$ profile [23]. However, the finite $\beta$ effects significantly influence the ITG instability in the weakly reversed magnetic shear configuration of the EAST ITB plasma.

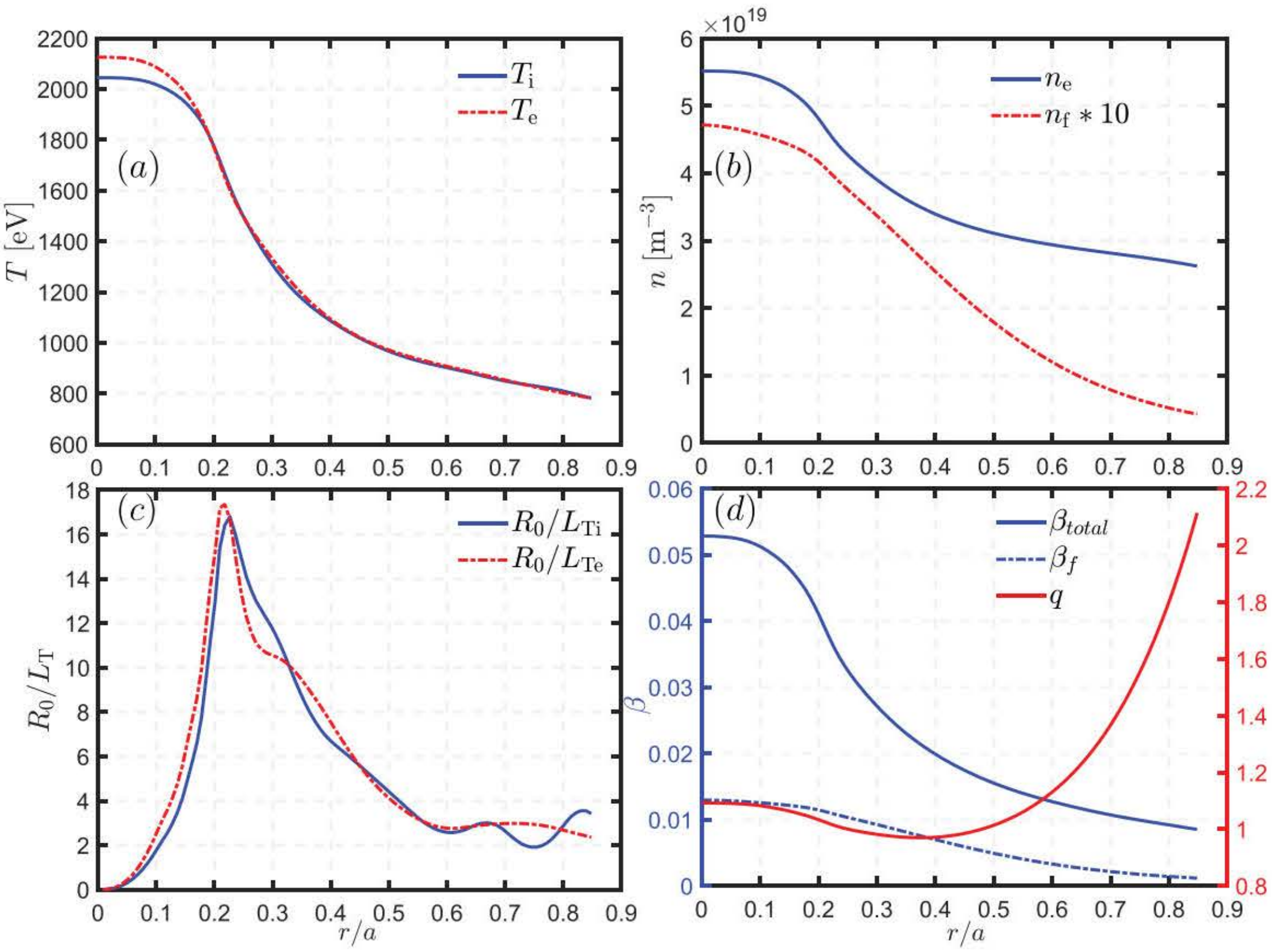

**Figure 1.** Plasma radial profiles in EAST discharge #93890: ($a$) displays temperatures $T$, ($b$) density $n$, ($c$) inverse temperature scale length $R_0/L_T$, and ($d$) the safety factor $q$ along with the ratio of plasma pressure to magnetic pressure $\beta = 8\pi nT/B_0^2$. Subscripts i, e, and f correspond to ions, electrons, and energetic ions (fast ions), respectively.

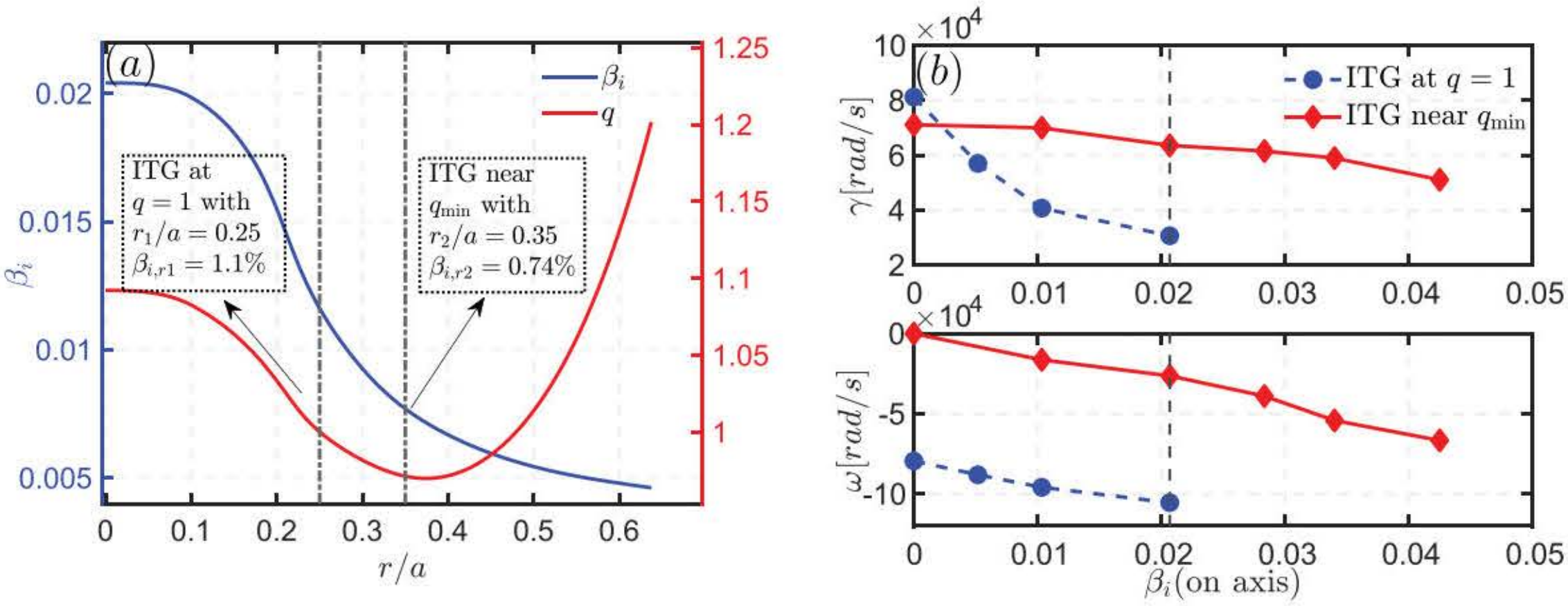

**Figure 2.** ($a$) shows the profiles of $\beta_i = 8\pi n_i T_i/B_0^2$ and $q$. The linear growth rate $\gamma$ (top panel) and real frequency $\omega$ (bottom panel) are displayed in ($b$) with torodial mode number $n = 20$. The blue circular curve represents the mode at the $q = 1$ surface, while the red diamond-shaped curve corresponds to the mode located near $q_{\rm min}$, each as a function of $\beta_i$ on axis. Different $\beta_i$ values are obtained by varying the electron density ($n_e = n_i$) while keeping the density gradient unchanged. The vertical dashed lines in ($b$) indicate the experimental value of $\beta_i$ on axis.

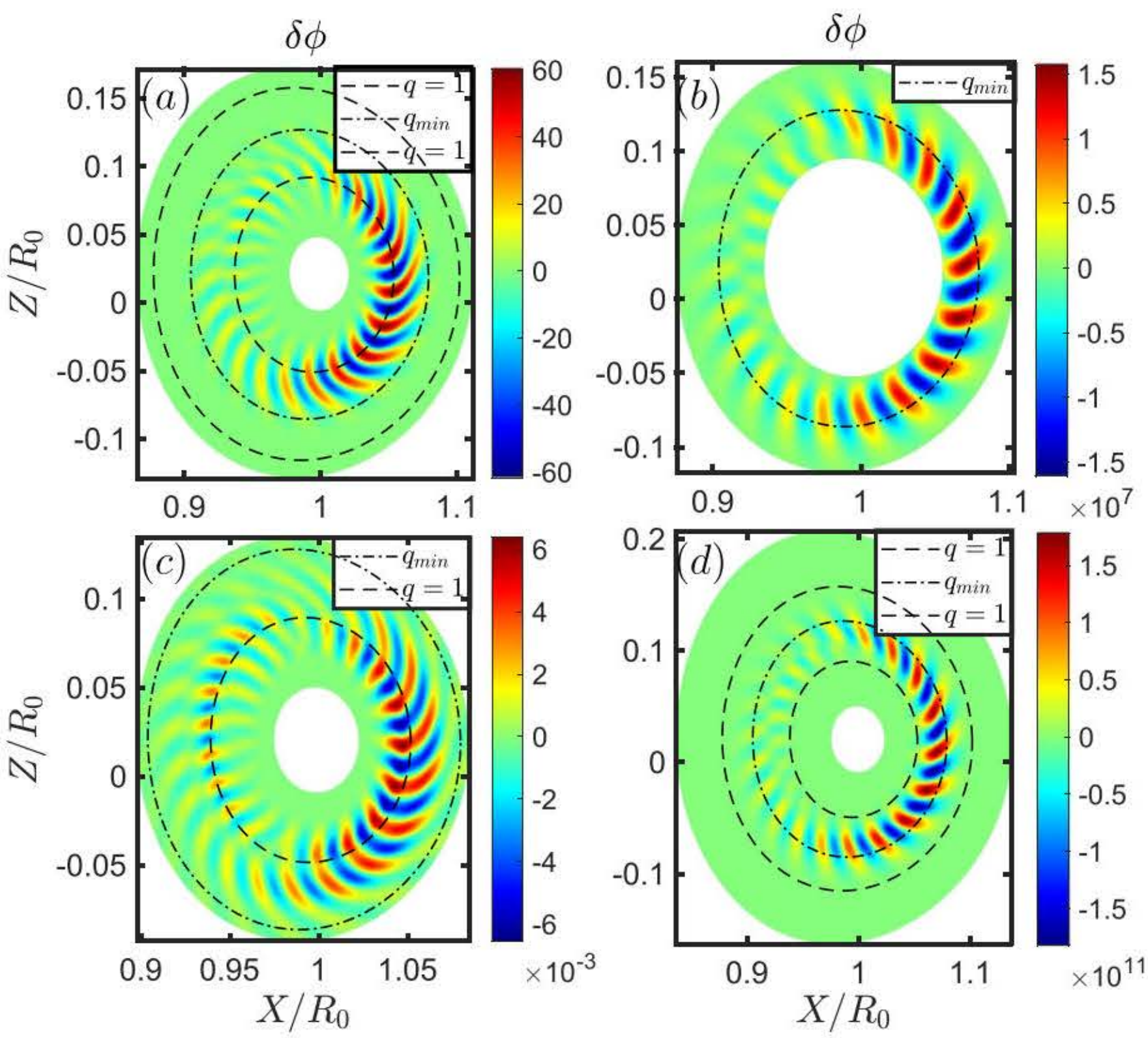

**Figure 3.** Mode structures of the electrostatic potential $\delta\phi$ with $n = 20$ obtained from electrostatic (top panel) and electromagnetic (bottom panel) linear simulations. Panels (*a*) and (*c*) show ITG modes at the $q = 1$ surface, while panels (*b*) and (*d*) show ITG mode near the $q_{\min}$ surface.

Next, we focus primarily on the ITG mode located near the $q_{\min}$ surface. The ITG mode with toroidal mode number $n = 25$ at this location is unstable at low $\beta_i$, with a small real frequency, as shown in figure 4(*a*). The vertical dashed lines in figure 4(*a*) indicate the experimental value of $\beta_{i,r_2}$ at the mode location. However, when $\beta_{i,r_2}$ exceeds approximately 3%, the KBM instability becomes unstable and its real frequency is significantly greater than that of the ITG mode. It is similar to the electromagnetic ITG and KBM instabilities observed in the CBC [23] and DIII-D tokamak pedestal studies [44]. Figures 5(*a*) and (*c*) illustrate that both ITG and KBM display ballooning structures in the electrostatic potential $\delta\phi$. The ITG perturbation exhibits a more pronounced ballooning angle, indicating a significant deviation from the out-midplane. In contrast, the KBM eigenmode structure is closer to the ideal ballooning mode. Figure 5(*b*) illustrates the ITG polarization, where the parallel electrostatic field $E_{\parallel}^{\mathrm{ES}} = -\mathbf{b}_0 \cdot \nabla\delta\phi$ is almost identical to the net parallel electric field $E_{\parallel}^{\mathrm{Net}} = -\mathbf{b}_0 \cdot \nabla\delta\phi - (1/c)\partial_t \delta A_{\parallel}$ indicating the quasi-electrostatic properties of the ITG mode. In contrast, as shown in figure 5(*d*) for the KBM, the amplitude of $E_{\parallel}^{\mathrm{Net}}$ is significantly smaller than that of $E_{\parallel}^{\mathrm{ES}}$ for all poloidal harmonics which leads to the predominantly Alfvénic polarization [53]. Finally, figure 4(*b*) presents the dispersion relation, showing the dependence of the linear growth rate and real frequency on the poloidal wavenumber (or equivalently, toroidal mode number $n$ ). The most unstable toroidal mode $n = 25$ corresponds to a poloidal wavenumber $k_\theta \approx 1.9\,\mathrm{cm}^{-1}$, close to values measured by poloidal correlation reflectometry (PCR) [54].

### 2.4. Nonlinear simulation results

In this subsection, global electromagnetic multi-$n$ nonlinear simulations are conducted to investigate the saturation and transport mechanisms of the ITG mode in the ITB region. All nonlinear electromagnetic simulations are performed at the experimental value of $\beta_i$. The kinetic electron model

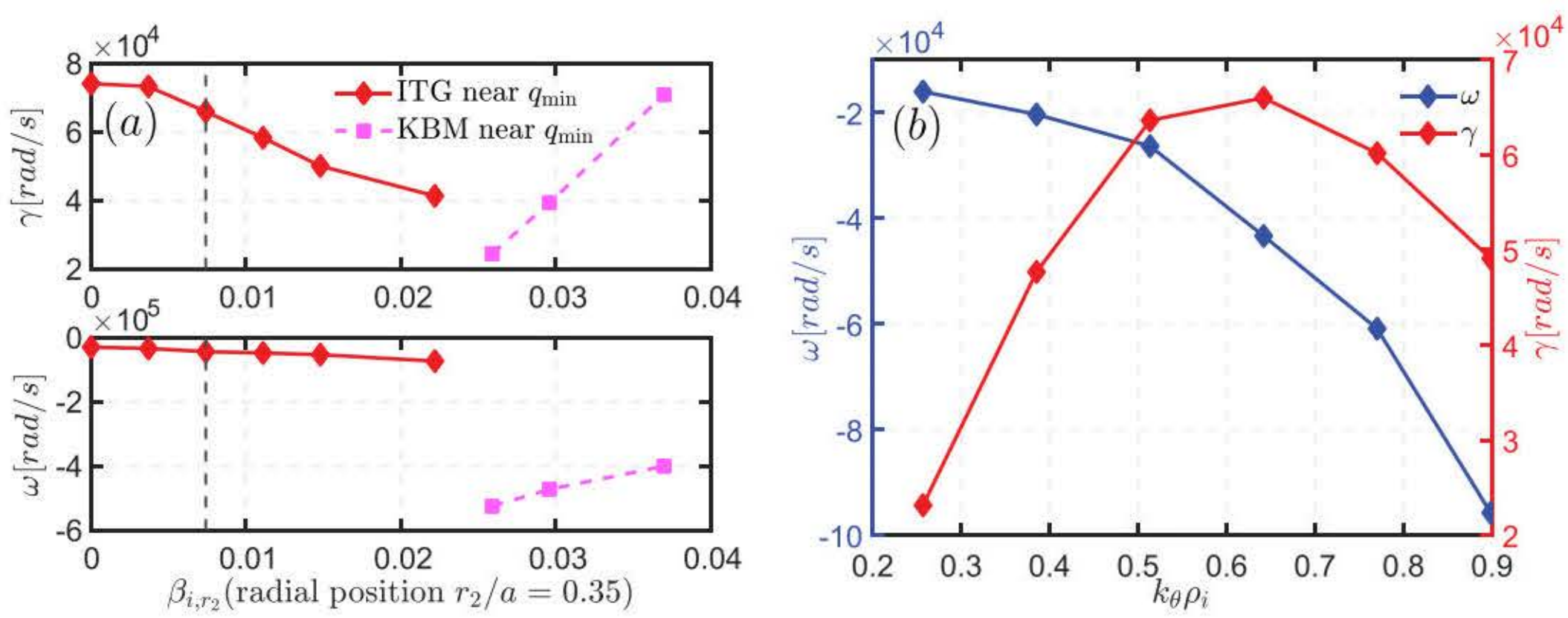

**Figure 4.** (*a*) The linear growth rate $\gamma$ (top panel) and real frequency $\omega$ (bottom panel) are shown for the mode located near $q_{\min}$ with $n = 25$, as a function of $\beta_{i,r_2}$ which is evaluated at mode location. (*b*) Dependence of the ITG mode near the $q_{\min}$ surface growth rate and real frequency on the poloidal wavelength $k_\theta \rho_i$ (corresponding to toroidal mode numbers $n = 10, 15, \ldots, 35$).

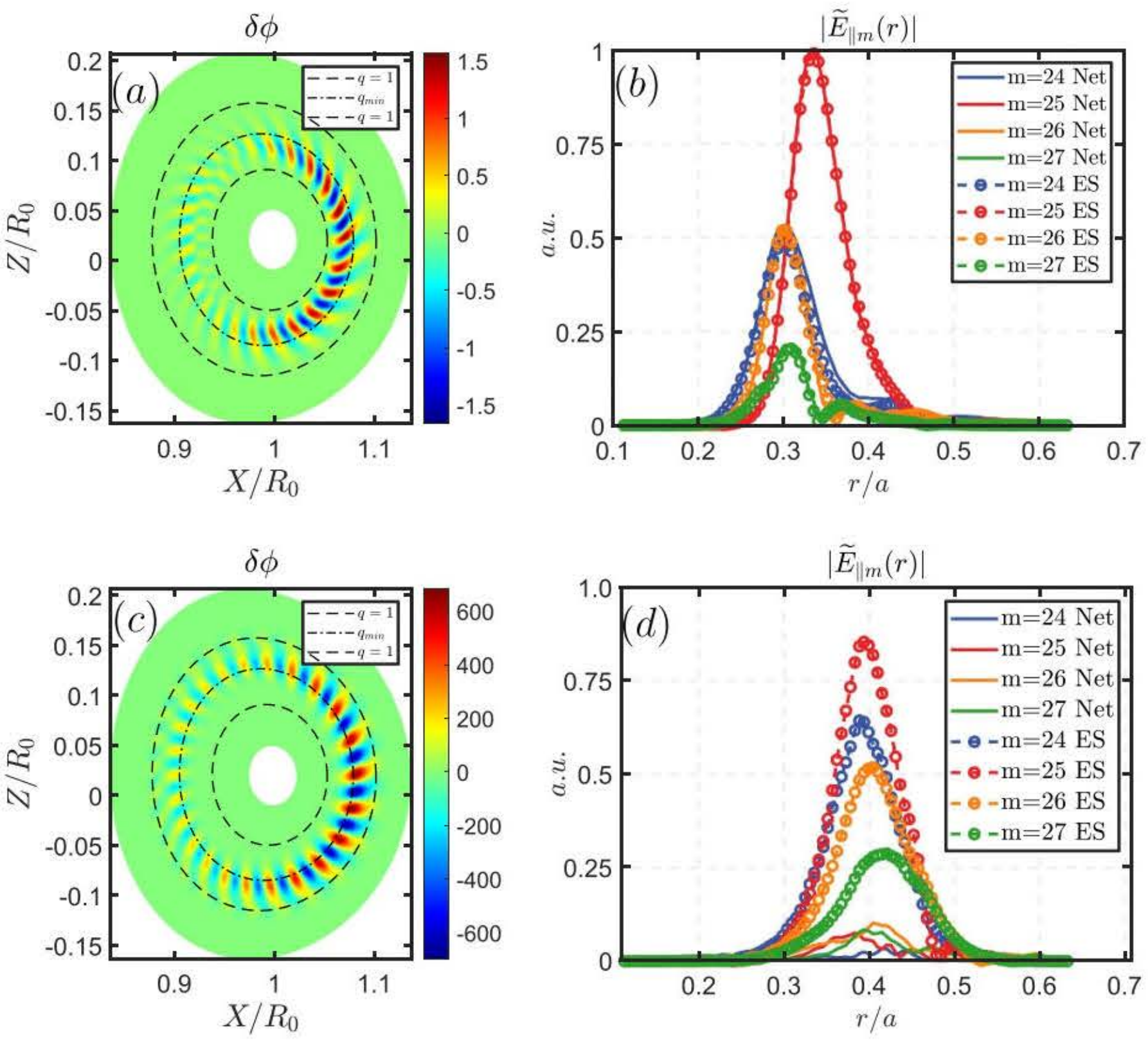

**Figure 5.** Mode structures of the $n = 25$ ITG mode (top panel) at $\beta_i = 1.11\%$ and the KBM mode (bottom panel) at $\beta_i = 3.7\%$: (*a*) and (*c*) show the poloidal contour plots of the $\delta\phi$, while (*b*) and (*d*) display the poloidal harmonics radial profiles of the parallel electric field $E_\parallel$. In (*b*) and (*d*), the solid lines with circles represent $E_\parallel^{\mathrm{ES}} = -\mathbf{b}_0 \cdot \nabla \delta\phi$, and the solid lines correspond to $E_\parallel^{\mathrm{Net}} = -\mathbf{b}_0 \cdot \nabla \delta\phi - (1/c)\partial_t \delta A_\parallel$.

is employed to consider the kinetic effects of electrons. The simulations include toroidal modes $n = 10, 11, 12, \ldots, 39$. Figures 6(*a*) and (*b*) illustrate the time evolution of the electrostatic potential $\delta\phi$ and parallel vector potential $\delta A_\parallel$ from the electromagnetic ITG simulation. Linearly most unstable ITG modes (e.g. $n = 25$) are firstly driven and dominate the early stage. The $\delta\phi_{n=25}$ and $\delta A_{\parallel n=25}$ exhibit exponential growth at the linear growth rate $\gamma_{n=25}$. In both the linear and intermediate regimes, $\delta\phi_{00}$ and $\delta A_{\parallel 00}$ grow exponentially at a growth rate $\gamma_{n,m=0} \approx 2\gamma_{n=25}$. This observation suggests that the zonal fields in ITG turbulence are passively generated via the so-called beat-driven process [55]. Near 0.25 ms, the ITG turbulence saturates and settles into a steady state nonlinear regime. At this stage, the amplitude of $\delta\phi_{n=25}$ is on the order of

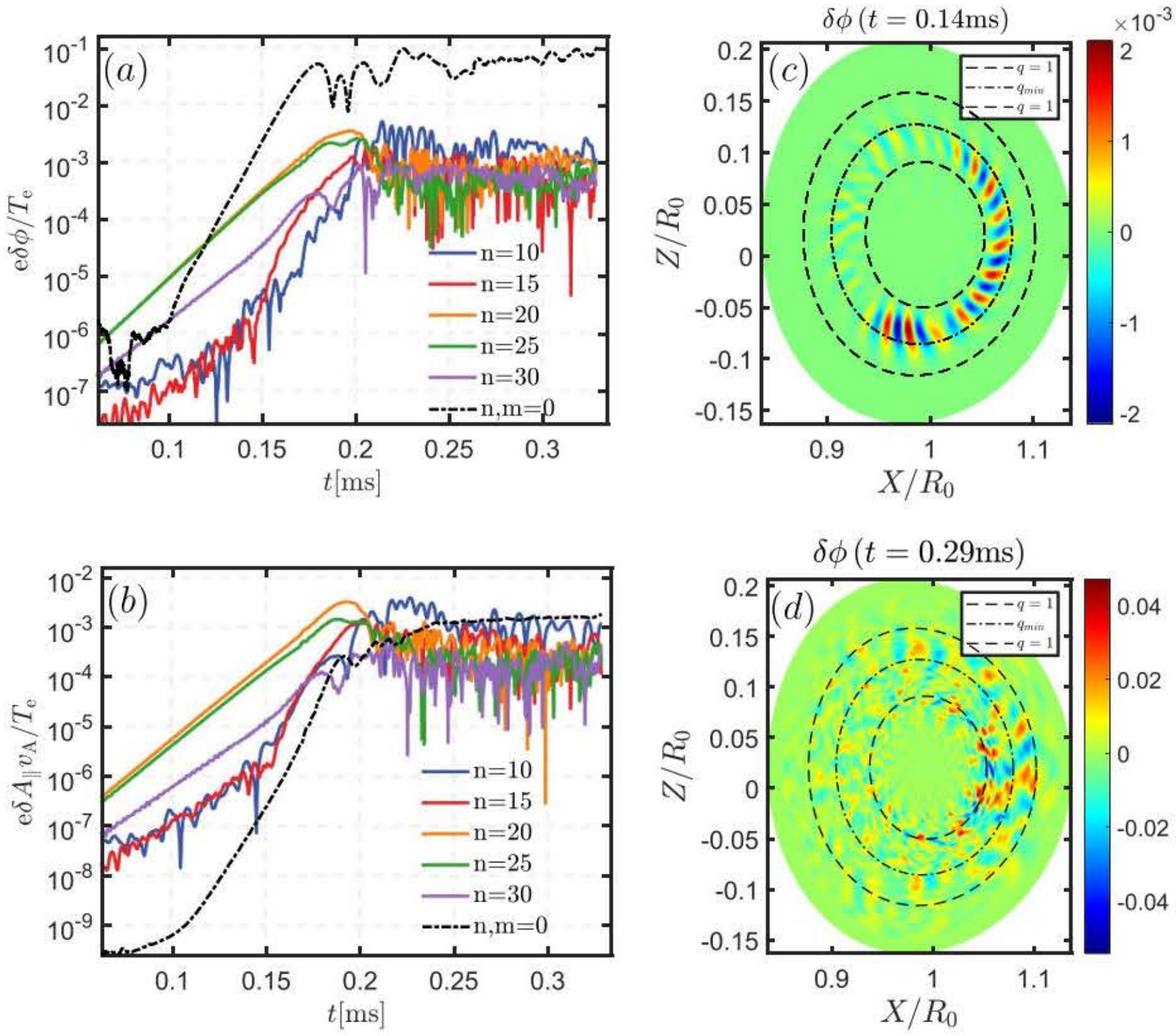

**Figure 6.** Time evolution of the perturbed electrostatic potential $\delta\phi$ and parallel vector potential $\delta A_{\|}$ for selected toroidal $n$ modes near the $q_{\min}$ flux surface from electromagnetic ITG simulations. The perturbed electrostatic potential and parallel vector potential are normalized as $e\delta\phi/T_e$ and $e\delta A_{\|}v_A/T_e$, respectively, where the Alfvén velocity is $v_A = B_0/\sqrt{4\pi n_{i0}m_i}$. The black dashed lines represent the zonal flow $\delta\phi_{00}$ and the zonal current $\delta A_{\|00}$, both shown as root-mean-square (rms) values averaged across the simulation domain. (*c*) and (*d*): Poloidal contour plots of the electrostatic potential $\delta\phi$ during the linear and nonlinear phases, respectively.

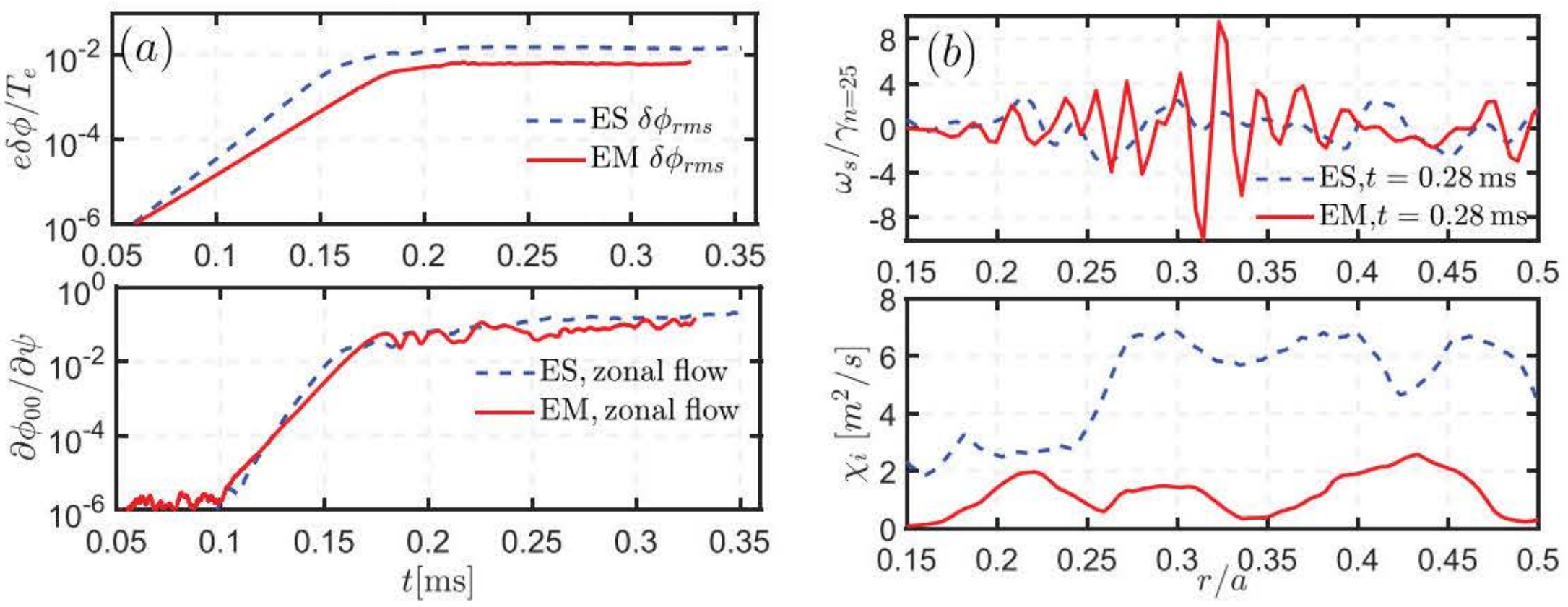

**Figure 7.** (*a*) Time evolution of the volume-averaged turbulence intensity $\delta\phi$ (top panel) and zonal flow (bottom panel) from electrostatic (dashed lines) and electromagnetic (solid lines) simulations. (*b*) The radial structures of the zonal flow shearing rate $\omega_s$ (top panel) and thermal ion heat conductivity (bottom panel) in the nonlinear phase for both electrostatic and electromagnetic cases, and the shearing rate normalized by the growth rate of the ITG mode with $n = 25$ near $q_{\min}$.

$10^{-3}$, while $\delta A_{\|}$ is on the order of $10^{-4}$. Figure 6(*c*) and (*d*) depict the mode structure of the perturbed electrostatic potential during the linear regime and the nonlinear saturation regime, respectively. From the contour plots, it is apparent that high $n$ modes (e.g. $n = 20$ and $n = 25$) dominate the electromagnetic ITG instability and are located between $q_{\min}$ and $q = 1$ in the linear phase. In contrast, low $n$ modes dominate the nonlinear phase and the turbulence spreads across the entire radial domain. The time history of the $\delta\phi$ and $\delta A_{\|}$ also show that once nonlinear saturation is reached, the amplitude of the low $n$ modes becomes larger than that of the high $n$ modes.

Figure 7(*a*) shows the time evolution of the volume-averaged perturbed electrostatic potential as well as the zonal flows for both electrostatic and electromagnetic ITG simulations. The turbulence intensity in the electromagnetic case is

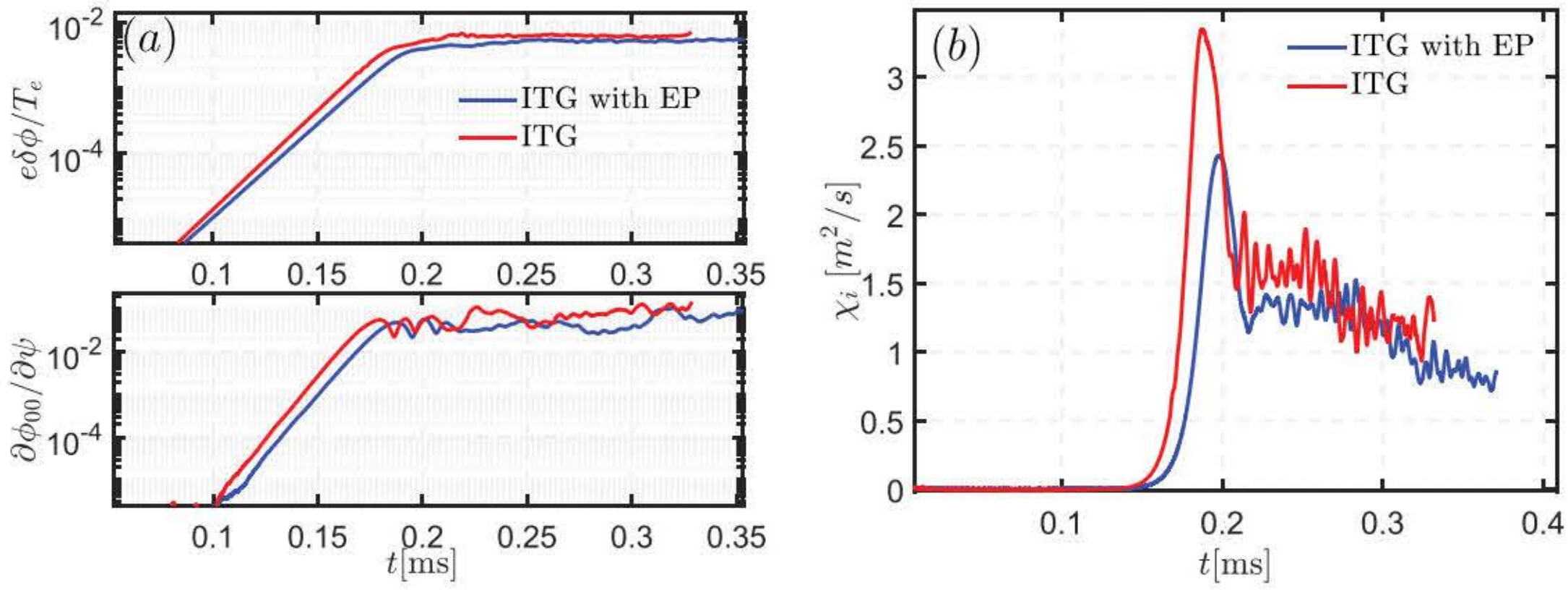

**Figure 8.** *(a)* Time evolution of the volume-averaged turbulence intensity $\delta\phi$ (top panel) and the zonal flow (bottom panel), and (*b*) the thermal ion heat conductivity, all obtained from electromagnetic ITG simulations with EPs (the blue line) and without EPs (the red line).

roughly three times smaller than that in the electrostatic case. The zonal flows amplitude is slightly smaller in the electromagnetic case. However, as shown in figure 7(*b*), the zonal flows shearing rate $\omega_s = -(RB_\theta)^2/B_0\,\partial^2\phi_{00}/\partial\psi^2$ [56, 57] is significantly larger in the electromagnetic case. Figure 7(*b*) also show the radial profile of thermal ion heat conductivity from both electrostatic and electromagnetic ITG simulations. The radial profiles of the thermal ion heat conductivity at the nonlinear saturation stage reveal that $\chi_i \gtrsim 1.5\,\mathrm{m^2\,s^{-1}}$ in the electromagnetic case, which is significantly smaller than $\chi_i \sim 6\,\mathrm{m^2\,s^{-1}}$ observed in the electrostatic case. It is observed that the ion heat conductivity is smaller in regions with higher shearing rate, indicating that the zonal flow shearing rate plays an important role in suppressing microturbulence. Notably, the radial region of reduced ion heat diffusivity $\chi_i$ and elevated normalized zonal flows shearing rate in figure 7(*b*) broadly coincides with the region of large $R_0/L_{Ti}$ shown in figure 1(*c*). This spatial correlation suggests that enhanced zonal flows shear can regulate turbulence even where the linear drive is strong. The thermal ion heat conductivity obtained from GTC agrees in magnitude with the results from the power balance analysis performed using the ONETWO code [37]. In summary, linear results indicate that the growth rate of the unstable mode in the electrostatic limit is approximately 1.25 times higher than that in the electromagnetic case. However, in the nonlinear phase of electromagnetic ITG, the transport coefficient is reduced by at least a factor of 4. The reduction is driven mainly by enhanced zonal flows shear that suppresses microturbulence during the nonlinear phase, with electromagnetic effects further reducing the linear growth rate.

Finally, the impact of EPs on electromagnetic ITG turbulence is investigated. As shown in figure 8, the presence of EPs leads to a slight reduction in both the ITG linear growth rate and the induced zonal flow. Moreover, at the nonlinear stage, the turbulence saturation level remains nearly unchanged, leading to only minor modifications in the ion heat conductivity. Overall, these findings indicate that EPs exert a stabilizing influence on ITG turbulence. This slight stabilization is likely attributed to the dilution effect of EPs ($n_f/n_e \lesssim 0.1$) and their finite $\beta$ effect [16, 58]. Evidently, the direct impact of EPs on ITG turbulence is minimal in our case. In the next step, we will explore the excitation of fishbone modes by EPs and their interaction with ITG turbulence through global nonlinear gyrokinetic cross-scale coupling simulations, aiming to further uncover the mechanisms underlying ITB formation.

## 3. Conclusions and discussions

In this paper, global gyrokinetic simulations are performed to investigate electromagnetic turbulence in the ITB region of an EAST tokamak discharge (#93890). GTC linear simulation results reveal the transition of the most unstable ITG mode at different radial positions due to the finite $\beta$ effects. Specifically, two dominant ITG modes at different radial positions: a higher frequency mode located at the $q = 1$ surface, which dominates in the electrostatic limit, and a lower frequency mode found near the $q_{\min}$ surface, which prevails in the electromagnetic regime. When electromagnetic effects are included, the ITG instability at the $q = 1$ surface is effectively suppressed, while the ITG mode near the $q_{\min}$ surface exhibits a smaller reduction in growth rate. Finite $\beta$ effects can more effectively stabilize the ITG mode at locations with a larger temperature gradient. It is therefore expected that electromagnetic stabilization effects may provide a negative feedback mechanism that regulates the turbulence driven by increasing temperature gradient during ITB formation.

Electromagnetic multi-$n$ nonlinear simulations primarily focus on the transport levels of turbulence. A comparison between electrostatic and electromagnetic turbulence demonstrates that including electromagnetic effects reduces the ion heat conductivity by at least a factor of 4. On one hand, electromagnetic effects play a important role in reducing the linear growth rate. On the other hand, although the zonal flows amplitude is smaller in the electromagnetic case compared to the electrostatic limit, the shearing effect of the zonal flows is significantly larger, resulting in a stronger suppression of microturbulence during the nonlinear phase. In this case, EPs exert a slight stabilizing influence on ITG turbulence, likely due to the dilution effect and finite $\beta$ contribution of EPs. In future work, we will study electromagnetic turbulent transport using gyrokinetic simulations under self-consistent magnetohydrodynamic equilibrium for each $\beta$.

## Acknowledgments

The authors acknowledge B. Zhang and T. Zhang for providing experimental data. This work was supported by the National MCF Energy R&D Program of China (Grant No. 2024YFE03050002), the Strategic Priority Research Program of Chinese Academy of Sciences (Grant Nos. XDB0790202 and XDB0500302), and National Natural Science Foundation of China (Grant Nos. 12025508 and 12275351). The numerical calculations in this paper were performed on the Hefei Advanced Computing Center, National Tianjin Supercomputing Center, and the Supercomputing Center of the University of Science and Technology of China.

## ORCID iDs

Yuehao Ma 0009-0007-8034-5402
Pengfei Liu 0000-0002-6739-3684
Jian Bao 0000-0002-2890-0700
Zhihong Lin 0000-0003-2007-8983
Huishan Cai 0000-0002-2500-3721